\documentclass[%
twocolumn,
reprint,
amsmath,amssymb,
aps,
]{revtex4-1}

\usepackage{graphicx}
\usepackage{dcolumn}
\usepackage{bm}
\usepackage{hyperref}

\hypersetup{
	colorlinks=true,
	linkcolor=blue,
	filecolor=magenta,      
	urlcolor=blue,
	citecolor=blue,
	anchorcolor=blue,
}

\begin{document}
	
	\preprint{APS/123-QED}
	
	\title{An Anomalous Reentrant 5/2 Quantum Hall Phase at Moderate Landau-Level-Mixing Strength}
	
	\author{Sudipto Das}
	\author{Sahana Das}%
	\author{Sudhansu S. Mandal}
	\affiliation{%
		Department of Physics, Indian Institute of Technology, Kharagpur, West Bengal 721302, India
	}%
	
	%
	%
	
	\date{\today}
	
	\begin{abstract}
		A successful probing of neutral Majorana mode in recent thermal Hall conductivity measurements opines in favor of the particle-hole symmetric Pfaffian (PH-Pf) topological order, contrasting the theoretical predictions of Pfaffian or anti-Pfaffian phases.	Here we report a reentrant anomalous quantized phase which is found to be gapped in the thermodynamic limit, distinct from the conventional Pfaffian, anti-Pfaffian, or PH-Pf phases, at an intermediate strength of Landau level mixing. Our proposed wave function consistent with the PH-Pf shift in spherical geometry rightly captures the topological order of this phase, as its overlap with the exact ground state is very high and it reproduces low-lying entanglement spectra. A unique topological order, irrespective of the flux shifts, found for this phase possibly corroborates the experimentally found topological order.
		
	\end{abstract}
	
	\maketitle
	
	The discovery \cite{willett_1987_ObservationEvendenominatorQuantum} of fractional quantum Hall effect in an even denominator filling factor 5/2 way back in 1987 had surprised the entire physics community. The first significant understanding of the possibility of this state had been put forward by Moore and Read (MR) through a unique proposal \cite{moore_1991_NonabelionsFractionalQuantum} of the Pfaffian (Pf) wave function, which was later interpreted \cite{read_2000_PairedStatesFermions} as a chiral p-wave pairing of the composite fermions \cite{jain_1989_CompositefermionApproachFractional,jain_2007_CompositeFermions} owing to their effective attractive interactions \cite{scarola_2000_CooperInstabilityComposite} in the second Landau level.
	Consequently, a flurry of new experimental techniques (See Ref.\cite{feldman_2021_FractionalChargeFractional} for a review)
	were developed for realizing exotic properties of this state such as quasiparticle charge, non-Abelian braiding statistics of the quasiparticles, and Majorana edge modes. On the other hand, subsequent proposals \cite{lee_2007_ParticleHoleSymmetryQuantum,levin_2007_ParticleHoleSymmetryPfaffian} of the anti-Pfaffian (A-Pf) which is topologically distinct from Pf yet degenerate for any two-body interaction makes the issue intriguing for understanding the true nature of the state. However, the Landau-level mixing (LLM), which is important for the second and higher Landau level quantum Hall states, generates three-body interaction \cite{bishara_2009_EffectLandauLevel,peterson_2013_MoreRealisticHamiltonians,simon_2013_LandauLevelMixing,sodemann_2013_LandauLevelMixing} that breaks \cite{peterson_2008_SpontaneousParticleHoleSymmetry} this degeneracy. A topologically distinct phase, namely, the particle-hole symmetric Pfaffian (PH-Pf)
	as the s-wave pairing of the Dirac composite fermions is also proposed \cite{son_2015_CompositeFermionDirac}, giving rise to yet another competing non-Abelian topological phase in the list for the same state. A PH-Pf-like instability in favor of the Dirac composite fermions with a greater mass arising due to LLM has
	also been proposed \cite{antonic_2018_PairedStatesParticlehole} in a gauge theory.

	The verdict of the numerical studies \cite{rezayi_2011_BreakingParticleHoleSymmetry,
		zaletel_2015_InfiniteDensityMatrix,
		pakrouski_2015_PhaseDiagramFractional,
		rezayi_2017_LandauLevelMixing,
		simon_2020_EnergeticsPfaffianAntiPfaffian} based on the LLM is possibly in favor of the A-Pf phase over the Pf phase.
	Both these topological phases and also the PH-Pf phase host the Majorana edge modes whose presence can be probed as half-integral thermal Hall conductance along the edge.  These phases are, however, distinguishable because the respective predicted values of thermal Hall conductances \cite{simon_2018_InterpretationThermalConductance,banerjee_2018_ObservationHalfintegerThermal} are $3/2$, $7/2$ and $5/2$ in the unit of  $G_0=\pi^2k_B^2T/(3h)$. In contrast to the theoretical expectation \cite{rezayi_2011_BreakingParticleHoleSymmetry,
		zaletel_2015_InfiniteDensityMatrix,
		pakrouski_2015_PhaseDiagramFractional,
		rezayi_2017_LandauLevelMixing,
		simon_2020_EnergeticsPfaffianAntiPfaffian,
		simon_2018_InterpretationThermalConductance},
the recent thermal Hall conductance and shot-noise measurements  \cite{banerjee_2018_ObservationHalfintegerThermal,dutta_2022_DistinguishingNonabelianTopological} are consistent with the PH-Pf phase. For reconciling this, several probable scenarios have been proposed in the literature such as non-equilibration  \cite{simon_2018_InterpretationThermalConductance}   of the thermal Majorana mode and subsequent proposals of partial-equilibration of anti-Pfaffian edge \cite{ma_2019_PartialEquilibrationInteger,simon_2020_PartialEquilibrationAntiPfaffian,asasi_2020_PartialEquilibrationAntiPfaffiana}, formation of puddles of Pf and A-Pf phases \cite{mross_2018_TheoryDisorderInducedHalfInteger,zhu_2019_DisorderDrivenTransitionFractional,zhu_2020_TopologicalInterfacePfaffian}, and stabilization of PH-Pf due to disorder \cite{lian_2018_TheoryDisorderedQuantum,mross_2018_TheoryDisorderInducedHalfInteger,wang_2018_TopologicalOrderDisorder,fulga_2020_TemperatureEnhancementThermal}.
However, no general consensus has yet been achieved and thus the 5/2 state remains enigmatic. Moreover, whereas the 5/2  state in GaAs is typically observed \cite{willett_1987_ObservationEvendenominatorQuantum,bishara_2009_EffectLandauLevel,peterson_2013_MoreRealisticHamiltonians,banerjee_2018_ObservationHalfintegerThermal,dutta_2022_DistinguishingNonabelianTopological,dean_2008_IntrinsicGapFractional,dean_2008_ContrastingBehaviorFractional,zhang_2010_FractionalQuantumHall,pan_2001_ExperimentalEvidenceSpinpolarized,pan_2014_CompetingQuantumHall,samkharadze_2017_ObservationAnomalousDensitydependent} in the range of $ 12$--$1 $ Tesla magnetic field which amounts to the LLM parameter $\kappa = 0.7$--$ 2.5 $, the theoretical studies \cite{rezayi_2011_BreakingParticleHoleSymmetry,zaletel_2015_InfiniteDensityMatrix,pakrouski_2015_PhaseDiagramFractional,rezayi_2017_LandauLevelMixing} have been performed for $\kappa \lesssim 1$ only. Also, a topological phase transition from a quantum Hall state to an unquantized state in the vicinity of $\kappa \sim 0.7$--1 followed by a hint of a new quantized phase for Pf shift has been found \cite{pakrouski_2015_PhaseDiagramFractional} by the numerical calculation of the lowest excitation energies and entanglement entropy.  However, the latter phase has not been explored further due to the lack of its understanding in terms of the MR wave function \cite{moore_1991_NonabelionsFractionalQuantum} $\Psi_{{\rm MR}}$ or its particle-hole conjugate \cite{lee_2007_ParticleHoleSymmetryQuantum,levin_2007_ParticleHoleSymmetryPfaffian} wave function $\Psi_{{\rm MR}}^{{\rm phc}}$.

Here we perform exact diagonalization of the Coulomb Hamiltonian corrected with LLM of strength $\kappa$
separately at Pf, A-Pf, and PH-Pf flux shifts for few-electron systems (up to $N=16$) in spherical geometry and determine overlaps of the exact ground states at different $\kappa$ values if the corresponding ground states are found at total angular momentum $L=0$. It amazingly shows (generic to all the shifts) clear segregation of two distinct quantum Hall phases formed at low and a moderate range of $\kappa$  separated by an unquantized (ground state at $L\neq 0$) regime in the vicinity of $\kappa \sim 0.7$. Whereas the overlaps of the ground states for intra-phase $\kappa$ values are nearly unity, it is negligible for inter-phase $\kappa$ values. A finite excitation gap for a pair of quasiparticle and quasihole in the thermodynamic limit for the phase at moderate-$\kappa$ suggests it to be a quantized phase.
The entanglement spectra (ES) for this quantized phase show an entanglement gap with
the number of edge state counting as 1-1-2-2 (up to the resolution obtained in finite systems).
Further, the ES for all the three flux shifts at this moderate-$\kappa$  {\em anomalous} phase ($\mathcal{A}$-phase) has a high degree of resemblance, signifying a unique topological order which is independent of these shifts.
We propose a trial wave function $\Psi_{\mathcal{A}}$ for this anomalous $\mathcal{A}$-phase. As its flux-shift matches with PH-Pf shift, we determine the overlap of it with the exact ground state at PH-Pf shift and found to be very high.
Moreover, the low-lying ES for $\Psi_{\mathcal{A}}$ is consistent with the same for the exact state and hence can be considered as representative for the true topological order of the  $\mathcal{A}$-phase.
We further analyze the topological properties of $\Psi_{\mathcal{A}}$ including the Majorana mode governing $2.5\,G_0$ thermal Hall conductance. Therefore, the $\mathcal{A}$-phase can possibly be attributed to the experimentally observed phase \cite{banerjee_2018_ObservationHalfintegerThermal,dutta_2022_DistinguishingNonabelianTopological} at moderate $\kappa$.

\begin{figure}[h]
	\centering
	\includegraphics[width=\linewidth]{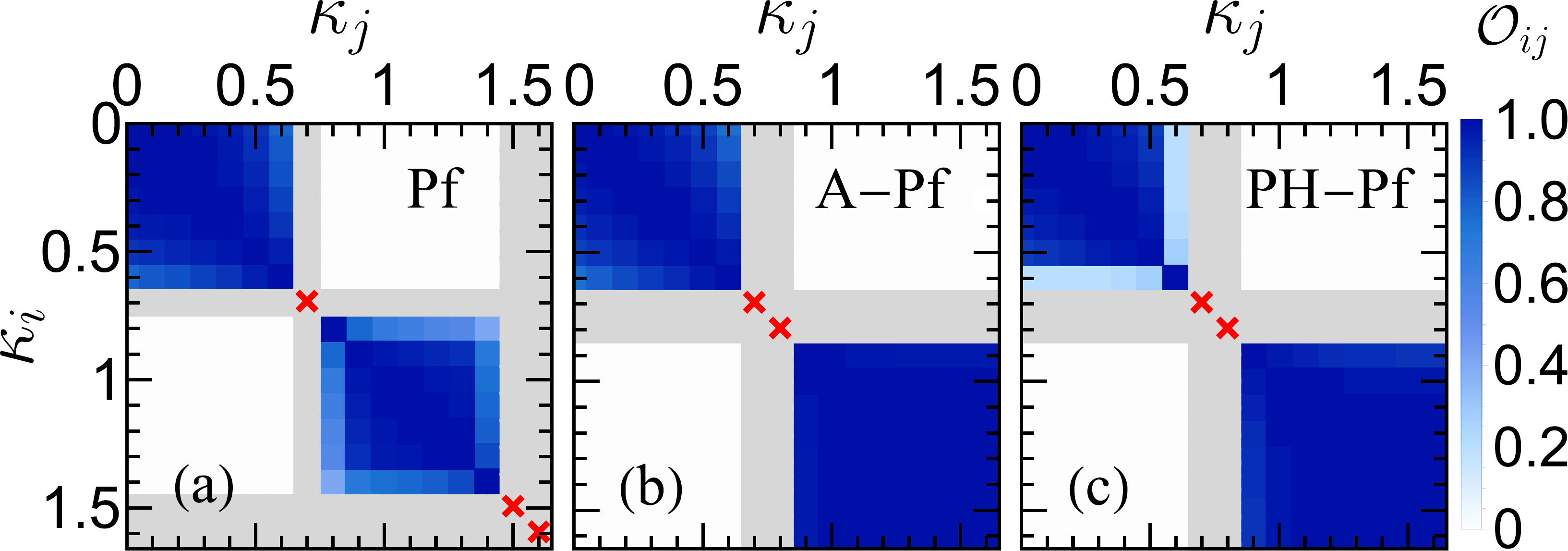}
	\caption{(color online) (a) Overlaps (shown as color map) of the exact ground states (found at $L=0$) of the Hamiltonian $\hat{H}_{\text{eff}}$ in Eq.~(1) with LLM-corrected pseudopotentials \cite{pseudopotential} at Pf flux shift ($N=14$ and $N_\phi=25$) for varying $\kappa$. Two topologically distinct phases are found as the overlaps of the ground states at different $\kappa$ values belonging to the same phase are closer to unity and two different phases are closer to zero. These two phases are intermediated by an unquantized regime (shown as crossed marks) of $\kappa$ where the ground state is not found at $L=0$. The gray zones indicate no overlap has been shown as one of the ground states corresponds to the unquantized regime.
		(b) Same as (a) but for A-Pf flux shift ($N=12$ and $N_\phi=25$).
		(c) Same as (a) but for PH-Pf flux shift ($N=14$ and $N_\phi=27$).
	}
	\label{fig.phase}
\end{figure}

The effective interaction  between spin-polarized electrons in the second Landau level with the consideration of LLM available in the literature \cite{bishara_2009_EffectLandauLevel,peterson_2013_MoreRealisticHamiltonians} as,
\begin{eqnarray}
	\hat{H}_{\text{eff}}(\kappa) &=&
	\sum_{m\, \text{odd}} \left[ V_m^{(2)} + \kappa \,\delta V_m^{(2)} \right] \sum_{i<j} \hat{P}_{ij}(m) \nonumber \\
	&&	+ \sum_{m\geqslant 3}
	\kappa \, V_m^{(3)} \sum_{i<j<k} \hat{P}_{ijk}(m) \, ,
	\label{eq.llmpot}
\end{eqnarray}
where  $ \hat{P}_{ij}(m) $ and $ \hat{P}_{ijk}(m) $ are two- and three-body projection operators respectively onto pairs or triplets of electrons with relative angular momentum $ m $. Here $V_m^{(2)}$ represents two-body Coulomb pseudo-potential in the second Landau level and $\delta V_m^{(2)}$ is its correction due to LLM and $V_m^{(3)}$ is three-body pseudo-potential arising due to LLM whose strength is defined by $\kappa =(e^2/\epsilon \ell_0)/\hbar \omega_c$ that is the ratio between Coulomb energy and cyclotron energy scales. We henceforth (unless mentioned otherwise) perform exact diagonalization of $\hat{H}_{\text{eff}}$  with pseudo-potentials \cite{pseudopotential} listed in Ref.~\onlinecite{peterson_2013_MoreRealisticHamiltonians}
for GaAs in spherical geometry for the Pf, A-Pf, and PH-Pf shifts, i.e., the respective number of flux quanta $N_\phi = 2N -3$, $N_\phi=2N+1$ and $N_\phi=2N-1$ with the shifts from $2N$, for different values of $\kappa$.
We determine overlaps of the exact ground states when found at $L=0$ for different values of $\kappa$, i.e., ${\cal O}_{ij}=\langle \Psi_{\text{gs}}(\kappa_i) \vert \Psi_{\text{gs}}(\kappa_j) \rangle $.

\begin{figure}[h]
	\centering
	\includegraphics[width=\linewidth]{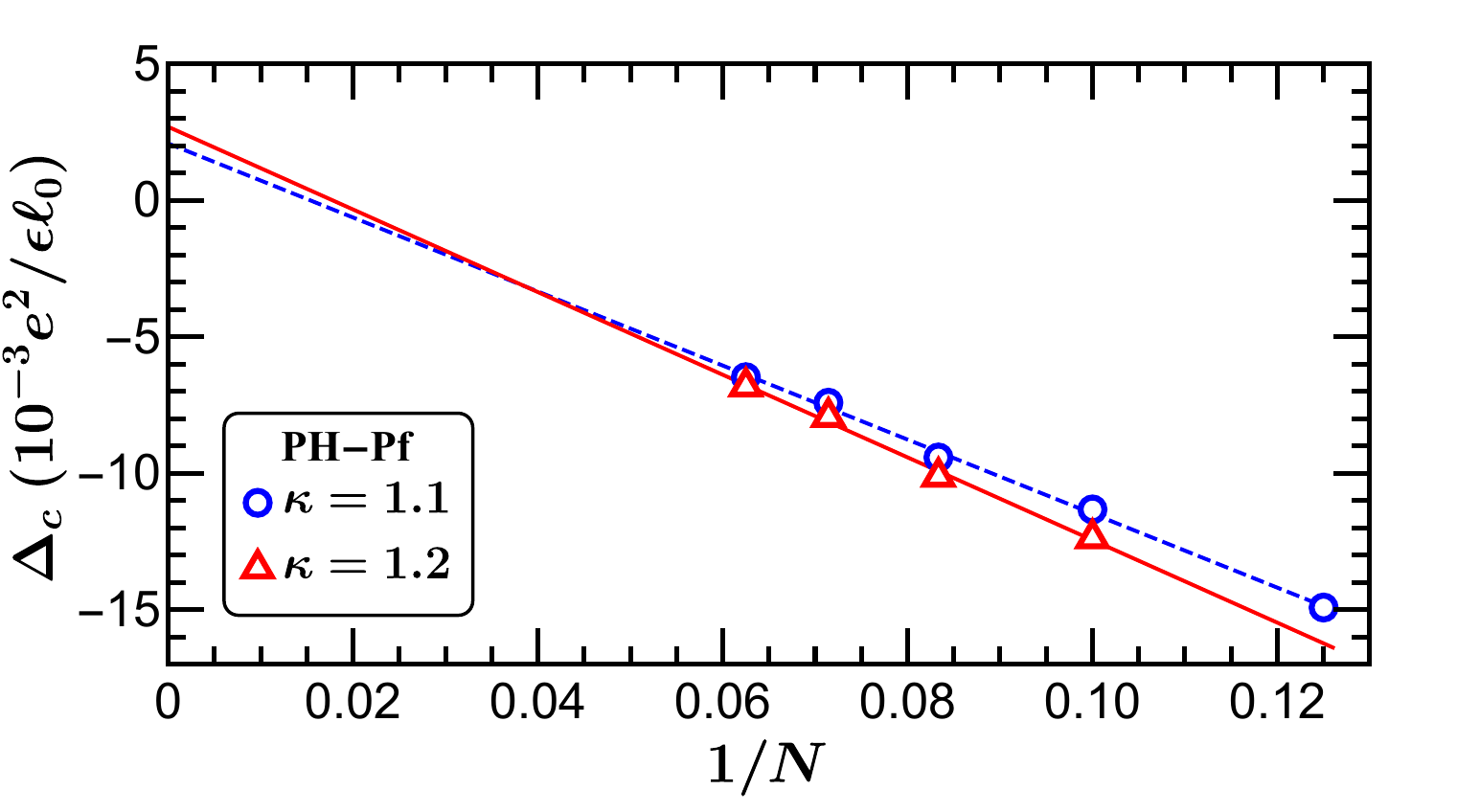}
	\caption{(color online) Scaling of Gap for a pair of quasiparticle-quasihole excitation at PH-Pf flux shift in the $ \mathcal{A} $-phase for $ \kappa=1.1 $. and 1.2 with $1/N$.}
	\label{fig.chargegap}
\end{figure}

In Fig.~\ref{fig.phase}(a)--(c), we show ${\cal O}_{ij}$ for Pf, A-Pf, and PH-Pf shifts respectively. The color mapping clearly shows two distinct topological phases separated by an unquantized phase in the vicinity of an intermediate value $\kappa_c \sim 0.7$ for all the three shifts. The transition between two quantum Hall phases is sharp (see supplemental material \cite{supplementary}) even for the finite width of the quantum wells, although the unquantized zone shrinks to zero at larger widths for finite size systems.  Whereas ${\cal O}_{ij}$ is nearly unity when both $\kappa_i$ and $\kappa_j$ belong to the same phase, it is vanishingly small when $\kappa_i$ and $\kappa_j$ belong to two different phases. We find (see supplemental material \cite{supplementary}) that the  $\mathcal{A}$-phase at moderate-$\kappa$ is present even in the absence of all $V_m^{(3)}$ but the presence \cite{pseudopotential} of  all $\delta V_m^{(2)}$.  Although a step-by-step addition of $V_3^{(3)}$, up to $V_3^{(8)}$, and finally up to $V_3^{(9)}$ reduces the range of the phase, it gets sharpened and the range reduction is compensated with the increase in $N$. Although the values of the pseudopotentials are estimated \cite{peterson_2013_MoreRealisticHamiltonians,rezayi_2017_LandauLevelMixing} perturbatively with perturbation parameter $\kappa$, because the $\mathcal{A}$-phase is robust against all these variations of pseudopotentials, we believe that the phase will be restored even for improved pseudopotentials at moderate-$\kappa$; the only quantitative change will be expected in terms of the change in the range of the phase in $\kappa$-space.

We further calculate the energy for creating a pair of quasiparticle and quasihole by taking an average \cite{rezayi_2021_StabilityParticleholePfaffian} of $E_{\rm qh}=E(N,2N)-E(N,2N-1)$ and $E_{\rm qp}=E(N,2N-2)-E(N,2N-1)$, {\it i.e.}, $\Delta_c = (E_{\rm qh}+E_{\rm qp})/2$, where $E(N,N_\Phi)$ denotes net ground state energy (after subtraction of the background energy of $N^2/\sqrt{2N_\Phi}$) of the system of $N$ electrons with $N_\Phi$ number of flux quanta. In Fig.~\ref{fig.chargegap}, we show $\Delta_c$ at different values of $N$ for $\kappa =1.1$ and $1.2$ belonging to $\mathcal{A}$-phase. Although each of these $\Delta_c$ is negative, their scaling with $1/N$ provides its thermodynamically extrapolated value $\Delta_c \sim 0.002$--$0.003\, e^2/(\epsilon \ell_0)$  which amounts to $\sim 200$--$300\, \text{mK}$ (in the same ballpark of the charged gap found in the experiments) for 5 Tesla magnetic field, where $\epsilon \simeq 13 $ is the dielectric constant of the host of the electron gas. The neutral excitations are also gapped (see supplemental material \cite{supplementary}). Therefore the $\mathcal{A}$-phase is incompressible and quantized.


\begin{figure}[h]
	\centering
	\includegraphics[width=\linewidth]{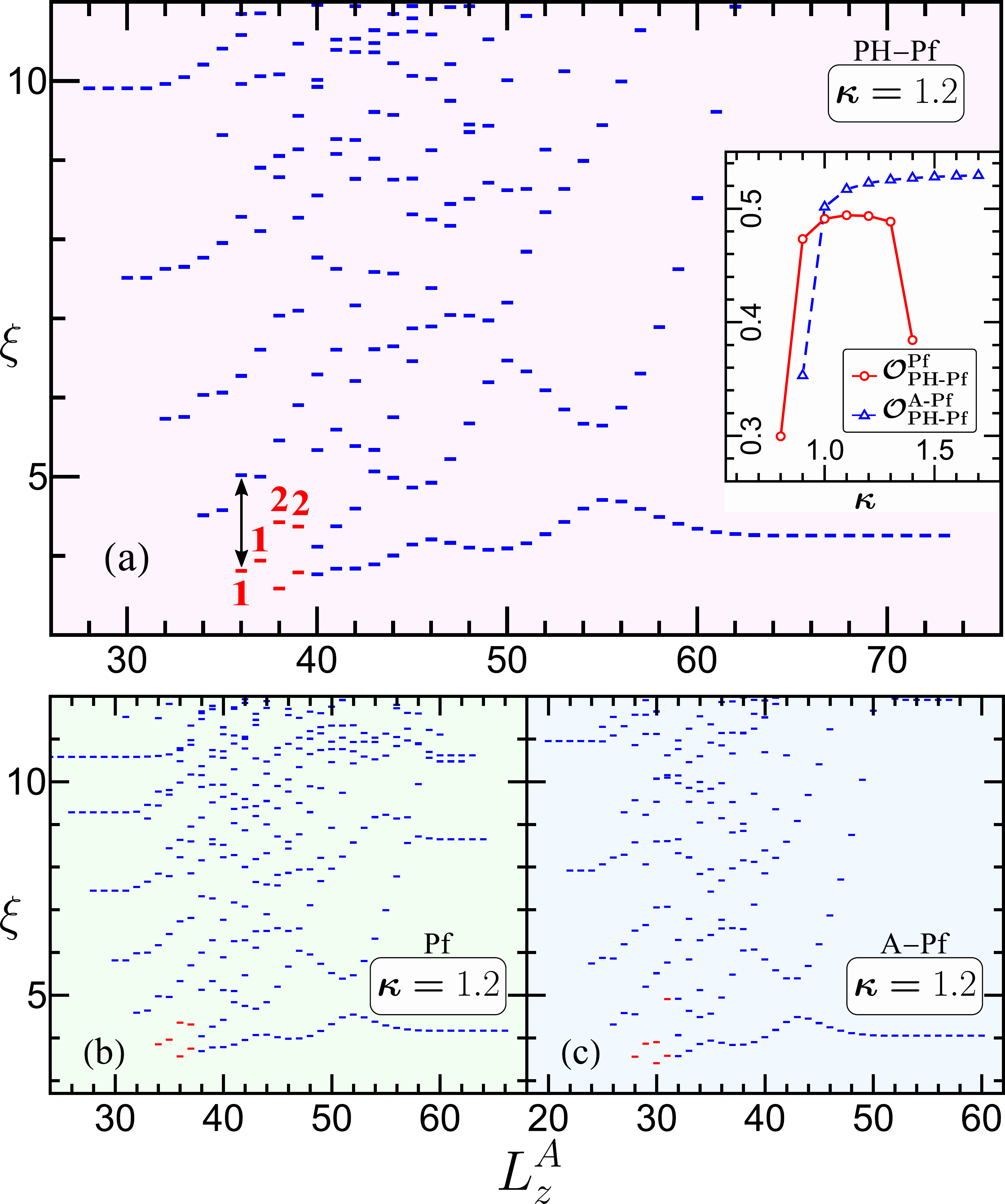}
	\caption{(color online)
		(a) ES for PH-Pf flux shift with $N_\phi=27$  at a $\kappa$ belonging to $ \mathcal{A} $-phase. $L_z^A$ represents the sum of the azimuthal components of angular momenta occupied by the particles in the A part of the partition \cite{supplementary}. An equal number of electrons ($N_A=N_B=7$) in both partitions are considered for computing the corresponding ES. Here $\xi$ represents entanglement energy in an arbitrary unit. The entanglement gap is shown by a line with an arrow-headed top and bottom, and the counting of edge states is marked. Inset: Overlaps of the ground state of PH-Pf flux ($N_\phi=27$) at $\kappa = 1.2$ with the ground states at Pf flux ($N_\phi=25$) and A-Pf flux ($N_\phi=29$) for different values of $\kappa$ in the range of $\mathcal{A}$-phase for $N=14$. The respective overlaps are $\mathcal{O}_{\rm PH-Pf}^{\rm Pf}$ (circles) and $\mathcal{O}_{\rm PH-Pf}^{\rm A-Pf}$ (triangles).
		(b) The ES for Pf flux shift with $N_\phi=25$, $N_A=N_B=7$, and $\kappa = 1.2$.
		(c) Same as (b) but for A-Pf flux shift with  $N_\phi=29$.
	}
	\label{fig.escomparision}
\end{figure}

We show (Fig.~\ref{fig.escomparision}) the ES \cite{li_2008_EntanglementSpectrumGeneralization,chandran_2011_BulkedgeCorrespondenceEntanglement,sterdyniak_2012_RealspaceEntanglementSpectrum}
for $\mathcal{A}$-phase for all three flux shifts (see supplemental material \cite{supplementary} for other phases). The low-lying ES for all the three shifts are broadly similar (Fig.~\ref{fig.escomparision}(a), (b) and (c)) in nature, suggesting its unique topological order. It does not belong to any of the conventional topological sectors, namely Pfaffian, anti-Pfaffian or particle-hole symmetric Pfaffian. The Hilbert space for PH-Pf (Pf) flux is a subspace of A-Pf (PH-Pf) flux (see supplemental material \cite{supplementary}) for a fixed $N$. The similar ES occurs because the ground-state at PH-Pf flux which is in between Pf and A-Pf fluxes for a fixed $N$ has
counter-intuitively (see supplemental material \cite{supplementary}) sizable overlap (inset Fig.~\ref{fig.escomparision}(a)) with the same for the latter two fluxes.   The ES is gapped and it displays the counting of edge states as 1-1-2-2-$\cdots$.

We propose a trial ground state wave function for the $\mathcal{A}$-phase of 5/2 state in the spherical geometry as
\begin{eqnarray}
	&&\Psi_{\mathcal{A}}(\{u_i,v_i\}) = \prod_{i<j}^N (u_iv_j -u_jv_i)  \nonumber \\
	&&\,\,\,\,\times {\cal S}\left[ \prod_{1\leqslant k,l\leqslant N/2} \left(u_kv_{N/2+l} - u_{N/2+l}v_k
	\right)^2\right]
	\label{wave_function}
\end{eqnarray}
where $u_j = \cos(\theta_j/2)e^{i\phi_j/2}$ and $v_j = \sin(\theta_j/2)e^{-i\phi_j/2}$ are the spherical spinors in terms of spherical angles $0 \leq \theta_j \leq \pi$ and $0 \leq \phi_j \leq 2\pi$, and ${\cal S}$ represents symmetrization in particle indices.
This wave function corresponds to $N_\phi = 2N-1$, i.e., PH-Pf flux shift {\it a la} the previously proposed PH-Pf wave function \cite{zucker_2016_StabilizationParticleHolePfaffian}, although the former doesn't have particle-hole symmetry. Owing to higher three-body interaction in the $\mathcal{A}$-phase, the system loses the particle-hole symmetry anyway.
The wave function $\Psi_{\mathcal{A}}$ in Eq.~(\ref{wave_function}) may be interpreted as two separate condensates of two-flavored composite bosons (electrons attached with one unit of flux quantum) with strong inter-flavored repulsive correlation. It further signifies that the bosonic wave function (ignoring the ubiquitous Jastrow factor required for Pauli exclusion principle for fermions) will not vanish even if the macroscopic $N/2$ bosons coincide. As per other known wave functions \cite{moore_1991_NonabelionsFractionalQuantum,read_1999_PairedQuantumHall}  with the possibility of coinciding two or more bosons supporting non-Abelian quasiparticles, the wave function $\Psi_{\mathcal{A}}$ in Eq.~(\ref{wave_function}) is likely to support non-Abelian quasiparticles.
The wave function in Eq. (\ref{wave_function}) may also be regarded as a fully antisymmetrized 113 Halperin wave function \cite{b.i_1983_TheoryQuantizedHall}.



\begin{figure}[h]
	\centering
	\includegraphics[width=\linewidth]{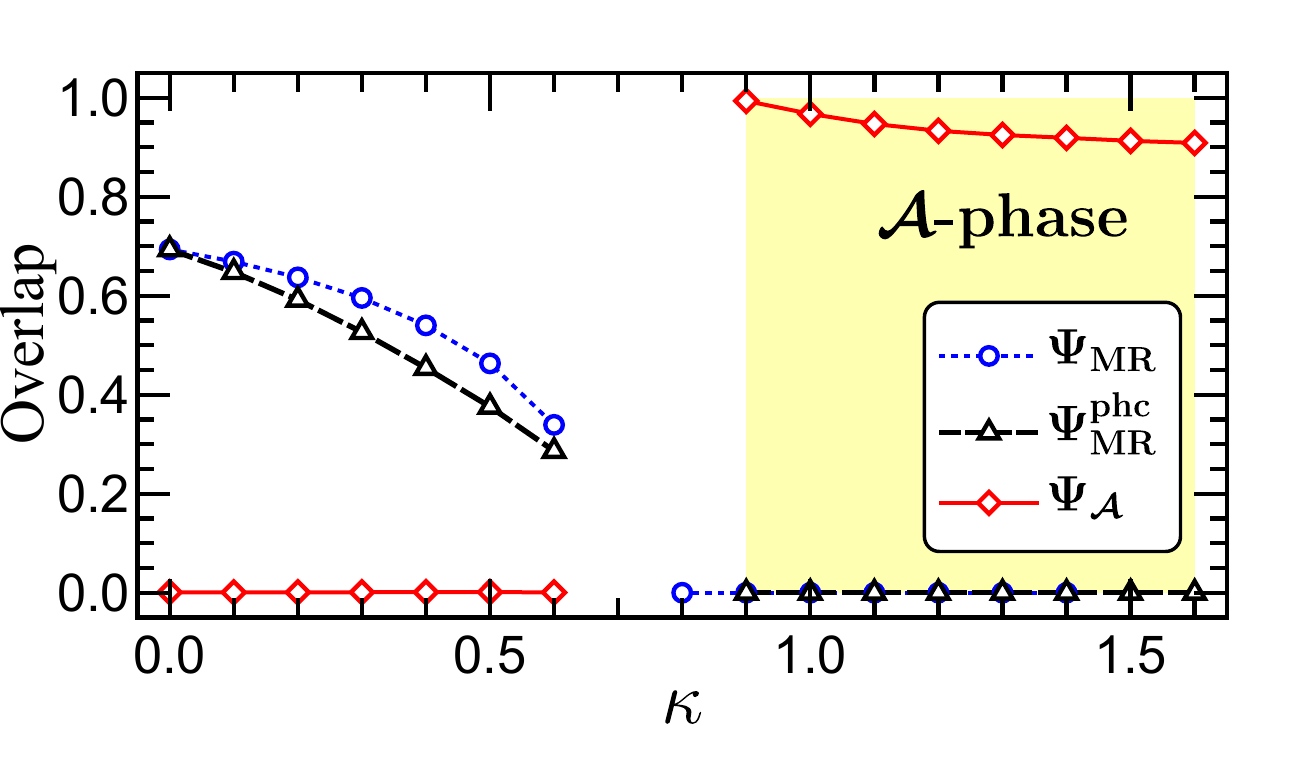}
	\caption{(color online) Overlaps of $\Psi_{\rm MR}$, $ \Psi_{\rm MR}^{\rm phc} $ (particle-hole conjugate of $\Psi_{\rm MR}$), and $\Psi_{\mathcal{A}}$ in Eq.~(\ref{wave_function}) with the corresponding exact ground states for $(N=14,\,N_\phi = 25)$, $(N=12,\, N_\phi = 25)$, and $(N=14,\, N_\phi = 27)$ respectively versus $\kappa$ at zero quantum well width. The unconnected zones refer to the unquantized regimes between two distinct topological quantized phases for each case.}
	\label{fig.overlap}
\end{figure}

\begin{figure}[h]
	\centering
	\includegraphics[width=\linewidth]{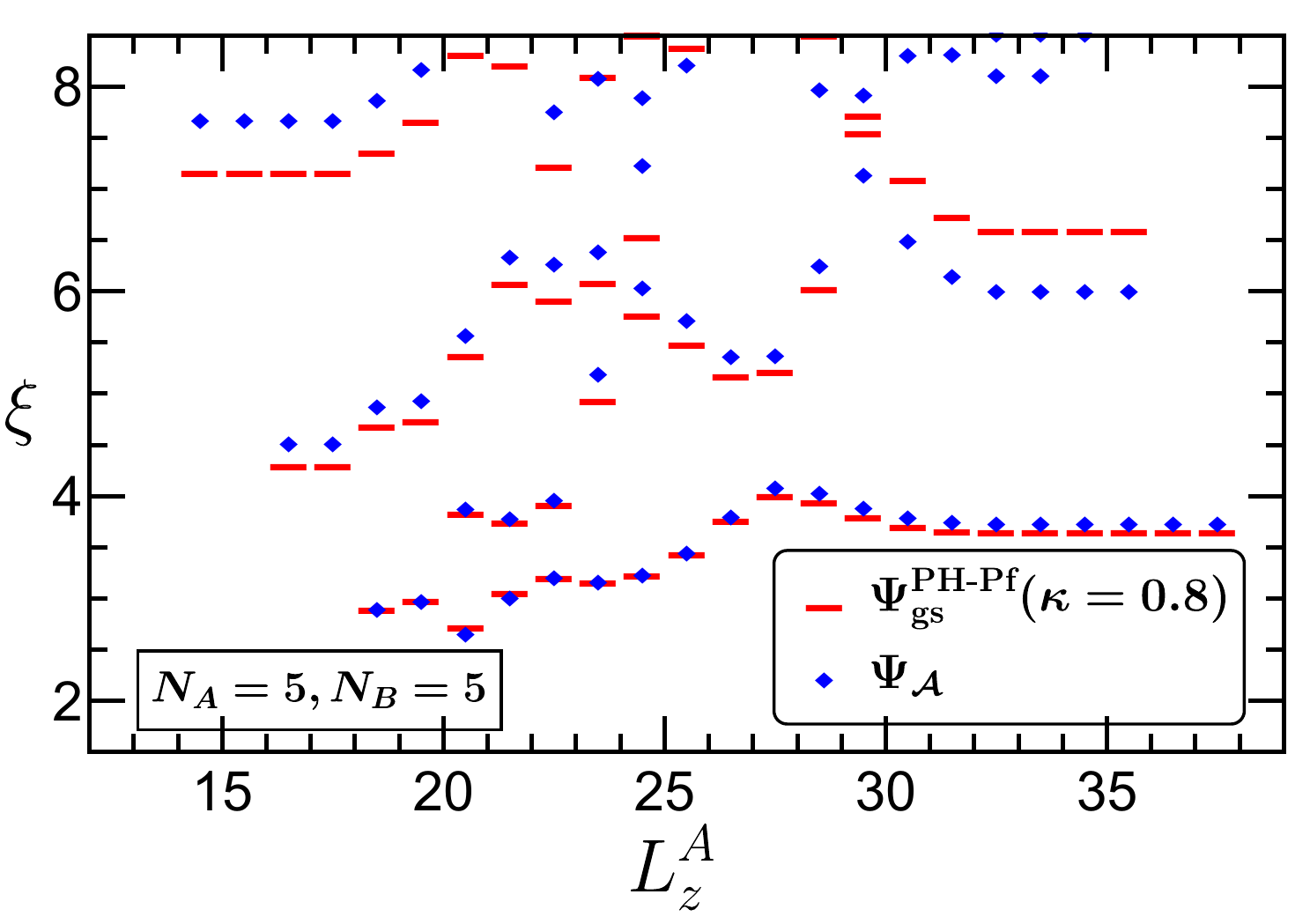}
	\caption{(color online) ES for $N=10$ and $N_\phi=19$ for the exact ground state (red dashed) and for $\Psi_{\mathcal{A}}$ (blue diamonds) at $\kappa =0.8$ belonging to quantized  $\mathcal{A}$-phase. Equal number of particles ($N_A=5,\, N_B=5$) have been considered for both the partitions for calculating the corresponding ES.}
	\label{fig.es10}
\end{figure}

In Fig.~\ref{fig.overlap}, we show the overlap of $\Psi_{\rm MR}$, $\Psi_{\rm MR}^{\rm phc}$, and $\Psi_{\mathcal{A}}$ with the corresponding exact ground states of $\hat{H}_{\text{eff}}$ at their respective flux-shifts. While the former two overlaps decrease with the increase of $\kappa$ and those are exceedingly low at moderate-$\kappa$ regime, the latter has, in contrast, very {\it high} overlap \cite{table_suplee} in the latter regime which coincides with the experimental regime of $\kappa = 0.8$--1.8.
We find, beyond this regime, another transition to an unquantized phase. However, as pointed out in Ref.~\onlinecite{pakrouski_2015_PhaseDiagramFractional}, the perturbatively obtained $\hat{H}_{\text{eff}}$ (Eq.~\ref{eq.llmpot}) may not be reliable at high $\kappa$ and thus such transition may defer to  higher $\kappa$.
Contrary to $\Psi_{\rm MR}$ and $\Psi_{\rm MR}^{\rm phc}$ wave functions, the wave function $\Psi_{\mathcal{A}}$ in Eq.~(\ref{wave_function}) with the exact ground states even at Pf and A-Pf fluxes in the $\mathcal{A}$-phase has a sizable overlap (see supplemental material \cite{supplementary}).
Therefore, the $ \mathcal{A} $-phase appears as independent of the flux shifts and is well characterized by the wave function $ \Psi_{\mathcal{A}} $.
%
%
Further, the low-lying ES corresponding to $\Psi_{\mathcal{A}}$ nicely resembles (see Fig.~\ref{fig.es10}) with the same for the exact ground state. Therefore, $\Psi_{\mathcal{A}}$ seems representing similar topological order as the exact ground state has for the 5/2 $\mathcal{A}$-phase.

The topological properties of the $\mathcal{A}$-phase can be extracted by exploiting the two-component structure of $\Psi_{\mathcal{A}}$. The corresponding low-energy effective Lagrangian density  \cite{wen_1992_ClassificationAbelianQuantum} is given by
\begin{equation}
	\mathcal{L} = -\frac{1}{4\pi}\epsilon^{\alpha\beta\gamma} \sum_{I,J=1}^2 K_{IJ} a_\alpha^I \partial_\beta a_\gamma^J - \frac{1}{2\pi}\epsilon^{\alpha\beta\gamma} \sum_{I=1}^2 t_I A_\alpha \partial_\beta a_\gamma^I
\end{equation}
with $K_{11}=K_{22}=1,\,K_{12}=K_{21}=3$ and $t_1=t_2=1$. Here $a_\alpha^1$ and $a_\alpha^2$ represent two components of Chern-Simons gauge fields, $A_\alpha$ is the external electromagnetic field, and $\epsilon^{\alpha\beta\gamma}$ is the antisymmetric Levi-Cevita tensor. Further introducing quasiparticle vector $l^T = (1,0)$ and spin vector $s^T = (1/2,1/2)$ and following Ref.~\onlinecite{wen_1992_ClassificationAbelianQuantum}, we find topological properties such as filling factor $\nu = t^T K^{-1} t =1/2$, quasiparticle charge $q= e\, l^T K^{-1}t=e/4$, topological shift $S = (2/\nu)t^T K^{-1}s =1$ relevant for Hall viscosity, and the ground state degeneracy $D = \vert {\rm Det}(K)\vert^g=8^g$ where $g$ is the genus of the geometry of the system. As the two eigenvalues of $K$ are opposite in sign, there will be one downstream charge mode with charge $q=e/4$ and one upstream neutral mode. The wave function $\Psi_{\mathcal{A}}$ has got hidden $\mathbb{Z}_2$ symmetry because the composite bosons are divided into two groups and up to $N/2$ composite bosons can occupy the same position. A similar hidden $\mathbb{Z}_2$ symmetry is present \cite{cappelli_2001_ParafermionHallStates} in the reformulated form of the Read-Rezayi \cite{read_1999_PairedQuantumHall} wave function for $5/2$ state. Because of the $\mathbb{Z}_2$ symmetry, one downstream neutral Majorana mode will also be present. Therefore, the net thermal Hall conductance becomes $2.5G_0$ as the fully filled lowest Landau level will provide the contribution of $2G_0$.

In this article, we show an {\it anomalous} topological 5/2 quantized fractional quantum Hall phase at a moderate strength of the Landau-level-mixing that is in the ballpark of the typical GaAs systems. We have proposed a wave function which turns out to be excellent for describing the ground state of this phase. Because the topological properties provided by this wave function are consistent with the enigmatic 5/2 state, we attribute the experimentally observed \cite{banerjee_2018_ObservationHalfintegerThermal} $2.5$ unit of thermal Hall conductance (not expected from the theoretical predictions for the conventional phases) in the system with $ \kappa \sim 1.1 $ to this phase. However, this assertion will be strengthened by the identification of
appropriate conformal field operators \cite{moore_1991_NonabelionsFractionalQuantum,hansson_2017_QuantumHallPhysics} whose correlation can
determine a wave function with a topological order
that should be adiabatically connected to our proposed wave function and at the same time explains
Majorana mode along with other charge modes for
the desired thermal Hall conductance.

Our work opens an avenue for exploring further characteristics of the $\mathcal{A}$-phase that is possibly the cause of the recent experimentally observed \cite{banerjee_2018_ObservationHalfintegerThermal,dutta_2022_DistinguishingNonabelianTopological} unusual value of thermal Hall conductance at 5/2 quantum Hall state. The role of Landau-level-mixing in the fractional quantum Hall states in graphene \cite{lin_2014_RecentExperimentalProgress,li_2017_EvendenominatorFractionalQuantum} and ZnO-based systems \cite{falson_2018_CascadePhaseTransitions,luo_2021_StabilityEvendenominatorFractional}, where it is stronger, should also be an interesting direction of future work; whether or not the predicted $\mathcal{A}$-phase occurs in such systems. Our results suggest a new approach for uncovering phases relevant in the experimental domain for other fractional quantum Hall states in the second Landau level, which are mostly enigmatic.

We thank Sutirtha Mukherjee for making useful comments on an early version of the manuscript.
We thank the developers of the DiagHam package for keeping it open access.
We acknowledge the Param Shakti (IIT Kharagpur) -- a National Supercomputing Mission, Government of India for providing computational resources.
S.S.M. is supported by Science and Engineering Research Board (Government of India) through Grant No. MTR/2019/000546.


%

\clearpage

\renewcommand{\figurename}{FIG. S\!\!}
\renewcommand{\tablename}{TABLE S\!}
\renewcommand*{\thesection}{S\arabic{section}}
\renewcommand\theequation{S\arabic{equation}}
\renewcommand{\bibnumfmt}[1]{[S#1]}

\setcounter{equation}{0}
\setcounter{figure}{0}


\begin{widetext}
	\centering
	\Large{\bf{ Supplemental Material for ``An Anomalous Reentrant 5/2 Quantum Hall Phase at Moderate Landau-Level-Mixing Strength"} }\\[1cm]
\end{widetext}

This supplemental material consists of six sections. In Section \ref{sec.finite width} , we determine phase diagrams by calculating overlaps between exact ground states at PH-Pf flux by varying LLM strength $\kappa$ for different widths of the quantum wells. In Section \ref{sec.Heff}, we obtain phase diagrams with step-by-step inclusion of three-body pseudopotentials of different relative angular momentum for zero width for confirming no adverse effect of any pseudopotentials that may change our conclusion in the main paper. In Section \ref{sec.neutral gap}, we show the neutral excitation gap with the variation of $\kappa$ at zero as well as finite widths of the quantum wells. In Section \ref{sec.ES EE}, we review the procedure of calculating ES and entanglement entropy, and show ES for all the three flux shifts for low-$\kappa$ phase and the subsequent unquantized phase. In Section \ref{sec.ovlp psiA difff flux}, we determine the overlap of $\Psi_{\mathcal{A}}$ with the exact ground states at three different fluxes for a fixed $N$. In Section \ref{sec.ovlp psiA PHPF}, we show the overlap of $\Psi_{\mathcal{A}}$ in Eq.~(2) with the exact ground states at PH-Pf flux for different values of $\kappa$ and $N$.

\begin{figure}[h]
	\centering
	\includegraphics[width=0.9\linewidth]{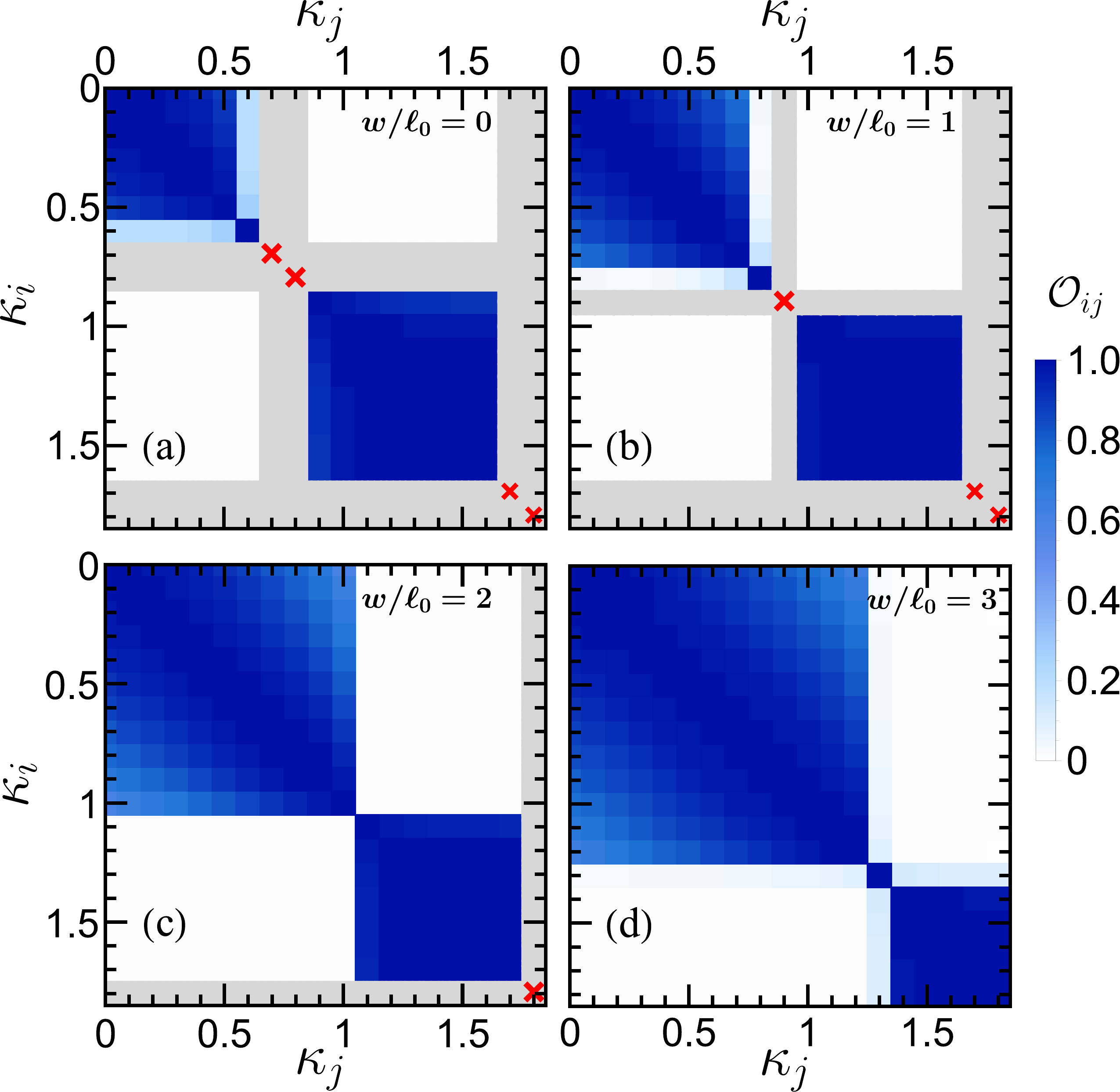}
	\caption{(color online)
		(a) The overlap matrix (color map) of exact ground states for different $ \kappa $ of the effective Hamiltonian $ \hat{H}_{\text{eff}} $ in Eq.~(1) at zero width, i.e. $ w/\ell_0 =0 $ at PH-Pf flux for $ N=14 $ particles.
		(b) Same as (a) but for $ w/\ell_0 = 1 $.
		(c) Same as (a) but for $ w/\ell_0 = 2 $.
		(d) Same as (a) but for $ w/\ell_0 = 3 $. The pseudopotentials used for this calculation is considered as tabulated in Ref.~11.
	}
	\label{fig:phpfphasewidthvaried}
\end{figure}


\section{\label{sec.finite width}Phase Diagrams for Finite Quantum Well Width}
In Fig.~S\ref{fig:phpfphasewidthvaried}(a)-(d), we show ${\cal O}_{ij}$
for the exact ground states of the effective Hamiltonian, $ \hat{H}_{\text{eff}} $ in Eq.~(1) corresponding to $ N=14 $ at PH-Pf flux considering finite width of $ w/\ell_0 = 0,1,2 $, and 3 respectively.
The two-body and three-body (up to $V_8^{(3)}$) pseudopotentials for the systems with finite widths have been taken from Ref.~11 for exact diagonalizations.
The $ \mathcal{A} $-phase appears at the moderate-$ \kappa $ and its transition from conventional low-$\kappa$ phase is sharp even when the effects of finite width is considered.  However, the regime in which $ \mathcal{A} $-phase appears shifts toward higher-$\kappa$ as width increases. Also, the unquantized phase disappears for higher widths in finite systems.

\begin{table}
	\begin{ruledtabular}
		\begin{tabular}{c|cccc}
			& \multicolumn{4}{@{}c@{}}{$ w/\ell_0 $}                                  \\
			\cline{2-5}
			\scalebox{1.2}{$\quad \kappa \quad $} & 0                                      & 1        & 2        & 3        \\
			\hline
			0                                     & 0.001031                               & 0.003069 & 0.000106 & 0.000258 \\
			0.1                                   & 0.000650                               & 0.003149 & 0.000194 & 0.000154 \\
			0.2                                   & 0.000884                               & 0.003246 & 0.000299 & 0.000040 \\
			0.3                                   & 0.001009                               & 0.003361 & 0.000423 & 0.000086 \\
			0.4                                   & 0.001159                               & 0.003494 & 0.000569 & 0.000221 \\
			0.5                                   & 0.001442                               & 0.003627 & 0.000735 & 0.000363 \\
			0.6                                   & 0.000570                               & 0.003728 & 0.000912 & 0.000509 \\
			0.7                                   & -                                      & 0.003767 & 0.001082 & 0.000654 \\
			0.8                                   & -                                      & 0.002774 & 0.001214 & 0.000798 \\
			0.9                                   & 0.993464                               & -        & 0.001244 & 0.000940 \\
			1                                     & 0.967333                               & 0.968478 & 0.001041 & 0.001085 \\
			1.1                                   & 0.946944                               & 0.942558 & 0.851365 & 0.001252 \\
			1.2                                   & 0.933147                               & 0.927919 & 0.90826  & 0.001524 \\
			1.3                                   & 0.924722                               & 0.918359 & 0.905918 & 0.071239 \\
			1.4                                   & 0.918953                               & 0.911577 & 0.901242 & 0.842247 \\
			1.5                                   & 0.912824                               & 0.906495 & 0.897057 & 0.865946 \\
			1.6                                   & 0.908678                               & 0.902535 & 0.893576 & 0.871898 \\
			1.7                                   & -                                      & -        & 0.890697 & 0.873432 \\
			1.8                                   & -                                      & -        & -        & 0.873493 \\
			1.9                                   & -                                      & -        & -        & -        \\
		\end{tabular}
	\end{ruledtabular}
	\caption{Overlap of $ \Psi_{\mathcal{A}} $ in Eq.~(2) with the exact ground state of Hamiltonian $ \hat{H}_{\text{eff}} $ in Eq.~(1) with variation of $ \kappa $ at different quantum well widths, $ w/\ell_0 $. Cells with dashed mark are unquantized points.}
	\label{tab.phpfwidthvaried}
\end{table}

In Table~S\ref{tab.phpfwidthvaried}, the overlaps of $ \Psi_{\mathcal{A}} $ with exact ground states of the effective Hamiltonian, $ \hat{H}_{\text{eff}} $ for widths $ w/\ell_0= 0, 1, 2, 3 $ are shown. Even when the effect of finite width is considered, $ \Psi_{\mathcal{A}} $ still posses very high overlap with the exact ground states in the $ \mathcal{A} $-phase.


\section{\label{sec.Heff}Phase Diagrams for Different Effective Pseudo-Potentials}
A detailed account of
the effective Hamiltonian by incorporating the effect of LLM   consisting of two-body and three-body pseudo-potentials (up to $ V_8^{(3)} $) corrections are available in Ref.~11. The importance of three-body pseudopotential $ V_9^{(3)} = - 0.0072\, e^2/(\epsilon \ell_0)$ has subsequently been shown [20] for switching over the topological order from Pf to A-Pf in the region of low-$\kappa$.  

\begin{figure}[h]
	\includegraphics[width=0.9\linewidth]{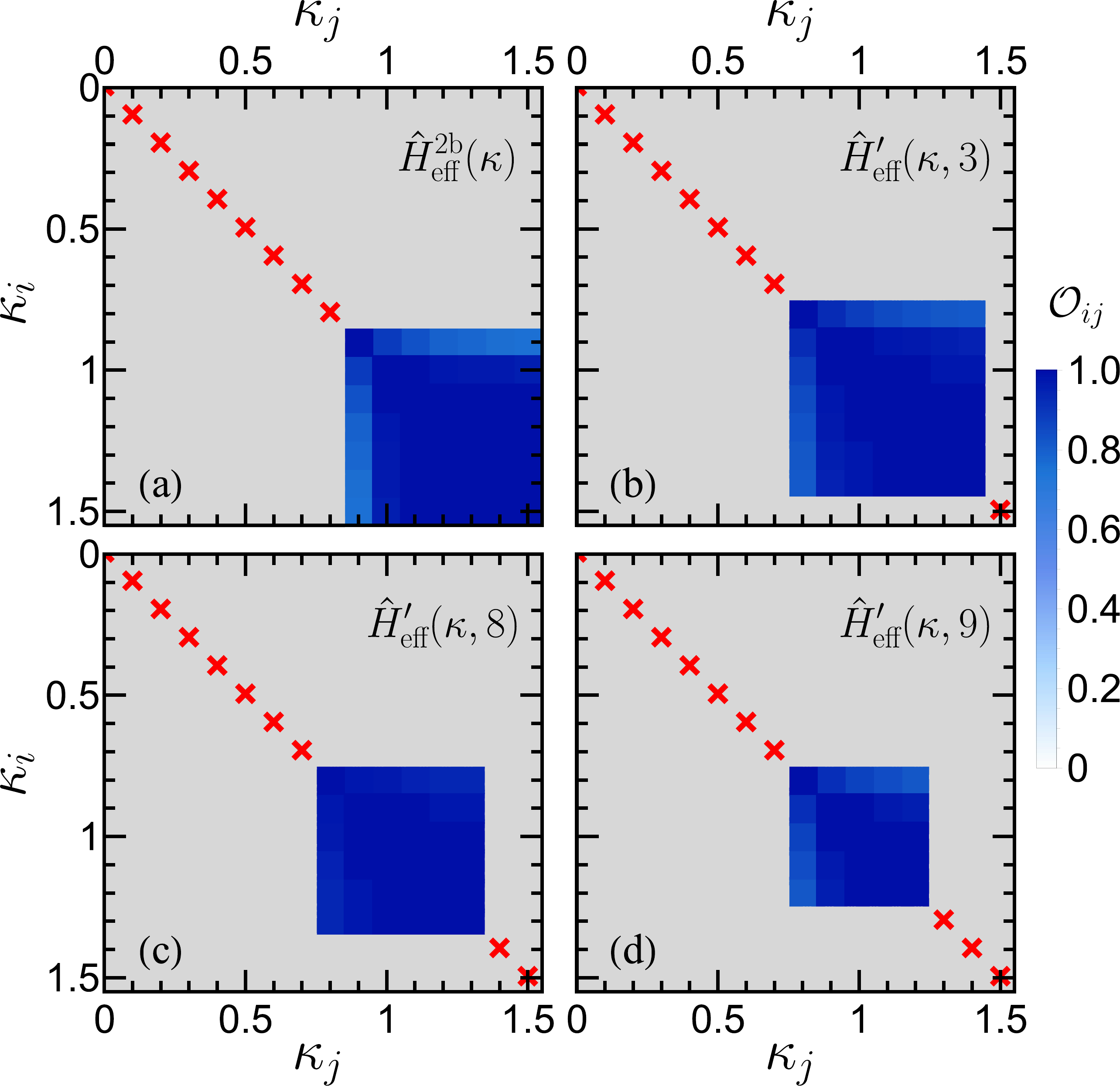}
	\caption{(color online)
		(a) The overlap matrix (shown as color map) of the exact ground states for different $ \kappa $ of the effective Hamiltonian, Eq.~(\ref{eq.heff2b})  incorporating 2-body pseudo-potential corrections [11] only at PH-Pf flux for $ N=10 $ particles. The unquantized points are marked with red cross and the unquantized regime is grayed out.
		(b) Same as (a) but for effective Hamiltonian in Eq.~(\ref{eq.heff}), $ \hat{H}_{\text{eff}}^{\prime}(\kappa, 3) $ i.e. 3-body $ V_3^{(3)} $ correction [11] only along with 2-body corrections.
		(c) Same as (a) but for effective Hamiltonian in Eq.~(\ref{eq.heff}), $ \hat{H}_{\text{eff}}^{\prime}(\kappa, 8) $ i.e. 3-body corrections upto $ V_8^{(3)} $ [11] along with 2-body corrections.
		(d) Same as (a) but for effective Hamiltonian in Eq.~(\ref{eq.heff}), $ \hat{H}_{\text{eff}}^{\prime}(\kappa, 9) $ i.e. 3-body corrections upto $ V_9^{(3)} $ [11, 20] along with 2-body corrections.
	}
	\label{fig.phasediag}
\end{figure}

\begin{table*}
	\begin{ruledtabular}
		\begin{tabular}{c|c|c|c|c|c}
			$ \quad \kappa \quad $
			& $ \langle \Psi_\mathcal{A} \mid \Psi_{\rm gs}^{\hat{H}_{\text{eff}}^{\rm 2b}(\kappa)} \rangle \quad $
			& $ \langle \Psi_\mathcal{A} + \Psi^{{\rm phc}}_\mathcal{A} \mid \Psi_{\rm gs}^{\hat{H}_{\text{eff}}^{\rm 2b}(\kappa)} \rangle \quad $
			& $ \langle \Psi_\mathcal{A} \mid  \Psi_{\rm gs}^{\hat{H}_{\text{eff}}^{\prime}(\kappa,3)}  \rangle  \quad $
			& $ \langle \Psi_\mathcal{A} \mid \Psi_{\rm gs}^{\hat{H}_{\text{eff}}^{\prime}(\kappa,8)}  \rangle  \quad $
			& $ \langle \Psi_\mathcal{A} \mid \Psi_{\rm gs}^{\hat{H}_{\text{eff}}^{\prime}(\kappa,9)}  \rangle  \quad $                                                                    \\
			\hline
			0   & -                                                                                                                                    & -       & -       & -       & -       \\ 
			0.1 & -                                                                                                                                    & -       & -       & -       & -       \\ 
			0.2 & -                                                                                                                                    & -       & -       & -       & -       \\ 
			0.3 & -                                                                                                                                    & -       & -       & -       & -       \\ 
			0.4 & -                                                                                                                                    & -       & -       & -       & -       \\ 
			0.5 & -                                                                                                                                    & -       & -       & -       & -       \\ 
			0.6 & -                                                                                                                                    & -       & -       & -       & -       \\ 
			0.7 & -                                                                                                                                    & -       & -       & -       & -       \\ 
			0.8 & -                                                                                                                                    & -       & 0.98490 & 0.99397 & 0.96590 \\ 
			0.9 & 0.70097                                                                                                                              & 0.99132 & 0.98005 & 0.96463 & 0.96501 \\ 
			1   & 0.66771                                                                                                                              & 0.94428 & 0.95079 & 0.94240 & 0.93139 \\ 
			1.1 & 0.63184                                                                                                                              & 0.89356 & 0.92946 & 0.92629 & 0.90654 \\ 
			1.2 & 0.61315                                                                                                                              & 0.86711 & 0.91315 & 0.91403 & 0.88753 \\ 
			1.3 & 0.60206                                                                                                                              & 0.85144 & 0.90005 & 0.90429 & -       \\ 
			1.4 & 0.59478                                                                                                                              & 0.84114 & 0.88920 & -       & -       \\ 
			1.5 & 0.58963                                                                                                                              & 0.83386 & -       & -       & -       \\ 
			1.6 & -                                                                                                                                    & -       & -       & -       & -       \\ 
		\end{tabular}
	\end{ruledtabular}
	\caption{Overlap of $ \Psi_{\mathcal{A}} $ with the exact ground states of different effective Hamiltonians for $ N=10 $ particle system at PH-Pf flux:  $\Psi_{\rm gs}^{\hat{H}_{\text{eff}}^{2b}(\kappa)}$ for $ \hat{H}_{\text{eff}}^{2b}(\kappa)$,  $\Psi_{\rm gs}^{\hat{H}_{\text{eff}}^{\prime}(\kappa,3)}$ for $ \hat{H}_{\text{eff}}^{\prime}(\kappa,3)$, $\Psi_{\rm gs}^{\hat{H}_{\text{eff}}^{\prime}(\kappa,8)}$ for $ \hat{H}_{\text{eff}}^{\prime}(\kappa,8)$, and $\Psi_{\rm gs}^{\hat{H}_{\text{eff}}^{\prime}(\kappa,9)}$ for $ \hat{H}_{\text{eff}}^{\prime}(\kappa,9)$.
		In the third column, overlap of the ground state  $\Psi_{\rm gs}^{\hat{H}_{\text{eff}}^{2b}(\kappa)} $ and $ 1:1 $ linear combination of $ \Psi_{\mathcal{A}} $ and it's particle-hole-conjugate, $ \Psi_{\mathcal{A}}^{{\rm phc}} $ is shown. The dashed marks indicate the exact ground states are not found at total angular momentum $L=0$. The conventional low-$\kappa$ phase is absent for $N=10$ at PH-Pf flux. }
	\label{tab.ovlp_Heff}
\end{table*}

In Fig.~S\ref{fig.phasediag}, we systematically show the effect of two-body and three-body pseudo-potential corrections to the phase diagram in ${\cal O}_{ij}$ for the system of zero width.
Incorporating only the two-body pseudo-potential corrections [11] 
to effective Coulomb pseudopotentials for the second Landau level  i.e,
\begin{eqnarray}
	\hat{H}_{\text{eff}}^{2b}(\kappa) &=&
	\sum_{m\, \text{odd}} \left[ V_m^{(2)} + \kappa \,\delta V_m^{(2)} \right] \sum_{i<j} \hat{P}_{ij}(m) \, ,
	\label{eq.heff2b}
\end{eqnarray}
the overlap matrix for the exact ground states with the variation of $ \kappa $ at PH-Pf flux for $ N=10 $ particles is shown in  Fig.~S\ref{fig.phasediag}(a). We note that there is no low-$\kappa$ quantized phase for 10 particle system at PH-Pf flux. Nevertheless, the $ \mathcal{A} $-phase emerges at moderate-$\kappa$ just for LLM two-body corrections. However, this phase will be particle-hole symmetric as any three-body pseudopotentials have not been considered in this case.
We then add the three-body pseudo-potentials up to relative angular momentum $ m =M$ along with two-body pseudo-potentials as described in the Hamiltonian,
\begin{eqnarray}
	\hat{H}_{\text{eff}}^{\prime}(\kappa, M) =
	\hat{H}_{\text{eff}}^{2b}(\kappa)
	+ \sum_{m=3}^M \kappa \, V_m^{(3)} \sum_{i<j<k} \hat{P}_{ijk}(m) \qquad
	\label{eq.heff}
\end{eqnarray}
In Fig.~S\ref{fig.phasediag}(b)--(c) we show the phase diagrams respectively for $M=3$, 8. The addition of more and more three-body pseudopotentials make the $\mathcal{A}$-phase sharper ($\mathcal{O}_{ij}$ is very close to unity for all $\kappa_i$ and $\kappa_j$). Although the range of $\mathcal{A}$ phase shrinks with more and more addition of pseudopotentials, it increases (Table~S\ref{tab.overlap}) with the increase in $N$.

As there are doubly degenerate  quantum states for three-body relative angular momentum $m=9$, namely, $\vert 0,3\rangle$ and $\vert 3,1\rangle$ where $m=2p+3q$ for a state $\vert p,q\rangle$ [13, S\onlinecite{laughlin_1983_AnomalousQuantumHall}], the corresponding Hamiltonian is given by [20],
\begin{eqnarray}
	H\vert_{m=9} &=& -0.0088 \mid 0,3 \rangle \langle 0,3 \mid
	+0.0033 \mid 3,1 \rangle \langle 3,1 \mid \quad  \nonumber\\
	&&+0.0007 \left[~ \mid 0,3 \rangle \langle 3,1 \mid +
	\mid 3,1 \rangle \langle 0,3 \mid ~\right]
	\label{eq.v9}
\end{eqnarray}
However, as pointed out in Ref.~20, Eq.~\eqref{eq.v9} can be well approximated by the dominant first term only with effective coefficient $ -0.0072 $,  i.e., $ V_9^{(3)} = -0.0072\, e^2/(\epsilon \ell_0) $. We then obtain phase diagram (Fig.~S\ref{fig.phasediag}(d)) by diagonalizing $\hat{H}_{\text{eff}}^{\prime}(\kappa, 9)$ in Eq.~\eqref{eq.heff}.
As per Fig.~S\ref{fig.phasediag}, the  $ \mathcal{A} $-phase is robust with respect to the degree of pseudo-potential corrections. Although the pseudo-potentials that have been used for obtaining phase diagrams in Fig.~1 of the main paper are obtained in Ref.~11 perturbatively which also possess good agreement with the other studies [12, 13],
this robustness against change in pseudopotentials indicates that the $\mathcal{A}$-phase is inevitable even if a better estimation of pseudopotentials is used in the moderate regime of $\kappa \sim 1.1$.


In Table~S\ref{tab.ovlp_Heff}, we show the overlap of  $ \Psi_{\mathcal{A}} $ in Eq.~(2) with the exact ground states for different Hamiltonian, $ \hat{H}_{\text{eff}}^{2b}(\kappa), \hat{H}_{\text{eff}}^{\prime}(\kappa, 3), \hat{H}_{\text{eff}}^{\prime}(\kappa, 8) $, and  $ \hat{H}_{\text{eff}}^{\prime}(\kappa, 9) $ for $ N=10 $ particles.
The overlaps for all the three Hamiltonian for which certain three-body potentials are considered are very high. In case of two-body only $\hat{H}_{\text{eff}}^{2b}(\kappa)$, the overlap is moderate, but that is expected because $\Psi_{\mathcal{A}}$ is not particle-hole symmetric whereas the exact ground state for  $ \hat{H}_{\text{eff}}^{2b}(\kappa)$, $\Psi_{\rm gs}^{\hat{H}_{\text{eff}}^{2b}(\kappa)}$ is particle-hole symmetric. However, $1:1$ linear combination of $\Psi_{\mathcal{A}}$ with its particle-hole conjugate $\Psi_{\mathcal{A}}^{\rm phc}$ provides high overlap (Table~S\ref{tab.ovlp_Heff}) with the corresponding exact ground state.



\section{\label{sec.neutral gap}Neutral Excitation Gap}
\begin{figure}[h]
	\centering
	\includegraphics[width=\linewidth]{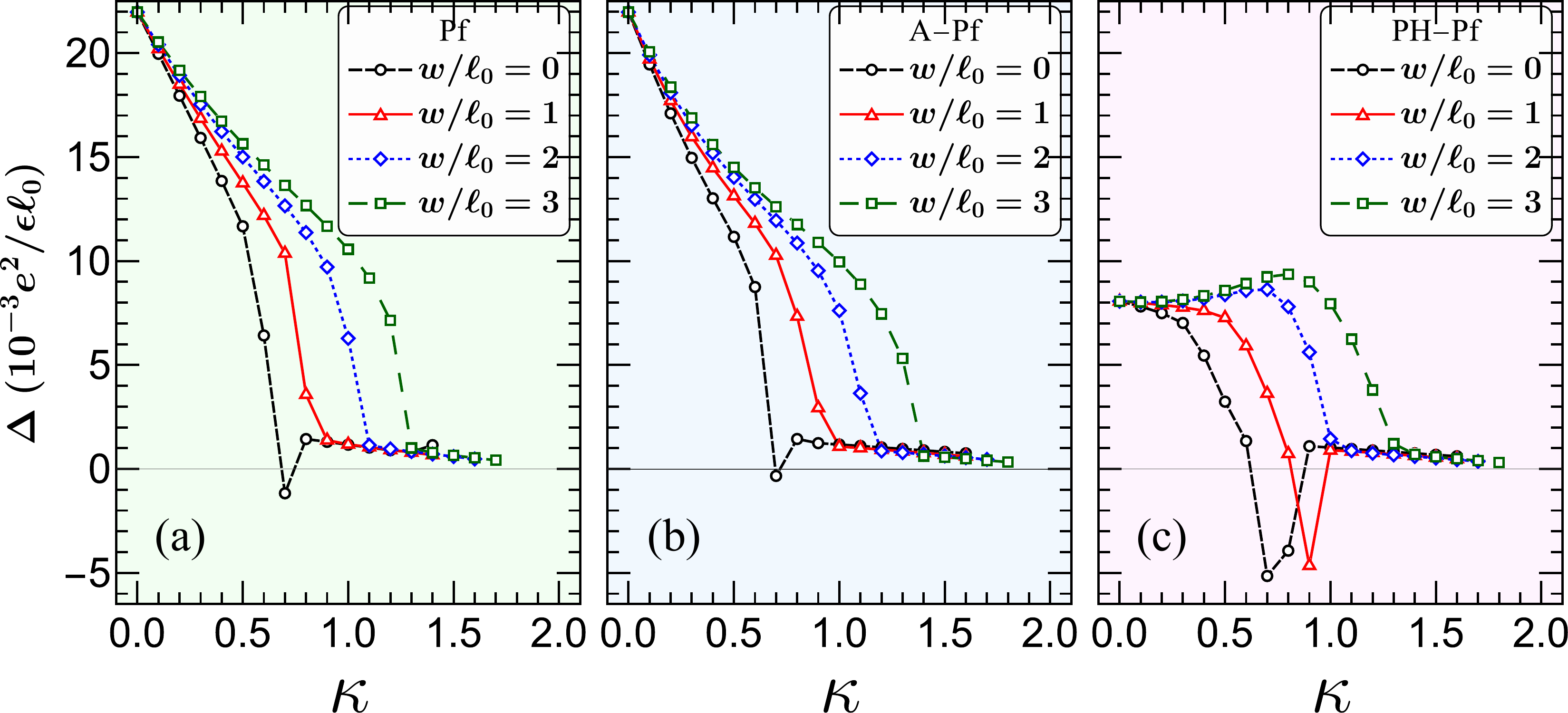}
	\caption{(color online)
		(a) The neutral excitation gap at $ N=14 $ and $ N_\phi=25 $ for quantum well widths $w = 0$, 1, 2, and 3 in the unit of $\ell_0$.
		(b) Same as (a) but for A-Pf shift with $N=12$ and $N_\phi=25$.
		(c) Same as (a) but for PH-Pf shift with $N=14$ and $N_\phi=27$.
	}
	\label{fig.n gap}
\end{figure}

The neutral excitation gap $ \Delta $, the energy difference of the minimum energy state at total angular momentum $ L = 0 $ and the minimum energy found amongst all other $ L $, is shown in Fig.~S\ref{fig.n gap}.

As $ \kappa $ increases,  $ \Delta $ decreases for all three fluxes and in the $ \mathcal{A} $-phase saturates to nearly $ 10^{-3} ~e^2/(\epsilon \ell_0) $. 
The neutral excitation gap, $ \Delta $ remains finite in the $ \mathcal{A} $-phase when the effect of finite width [11] is also included. Although when the finite width is considered, the states don't go through an unquantized phase (ground state at $ L \neq 0 $ thus negative $ \Delta $) which might be restored through further improvements of the LLM pseudo-potential corrections; however, the occurrence of the $ \mathcal{A} $-phase can not be hindered through such small perturbations.


\section{\label{sec.ES EE}Entanglement entropy and entanglement spectra}

The topological properties of a system can be unfolded with the study of quantum entanglement.
Usually, the entanglement of a quantum state is quantified by the entanglement entropy (EE) [S\onlinecite{zozulya_2007_BipartiteEntanglementEntropy}, S\onlinecite{haque_2007_EntanglementEntropyFermionic}], called the von Neumann entropy.
However, by the advent of a method of finding entanglement spectra (ES) of an FQHE state by Li and Haldane [43] and later by others [44, 45], one obtains more topological information of a quantum state in comparison to EE which is an integrated number.
To get the information about the entanglement of the state through EE and ES, the whole system is partitioned into two blocks, and thus the Fock space of the full Hamiltonian $\mathcal{H}$ is divided into two parts: $\mathcal{H} = \mathcal{H}_A \otimes \mathcal{H}_B$, where $\mathcal{H}_A$ and $\mathcal{H}_B$ respectively represent the Hamiltonian of parts \textit{A} and \textit{B}.
Here we adopt the method in Ref. 44 by making
an orbital partition for the particles on a sphere whose center possesses magnetic monopole charge $Q$ for generating $2Q$ number of magnetic flux.

\begin{figure}[t]
	\includegraphics[width=\linewidth]{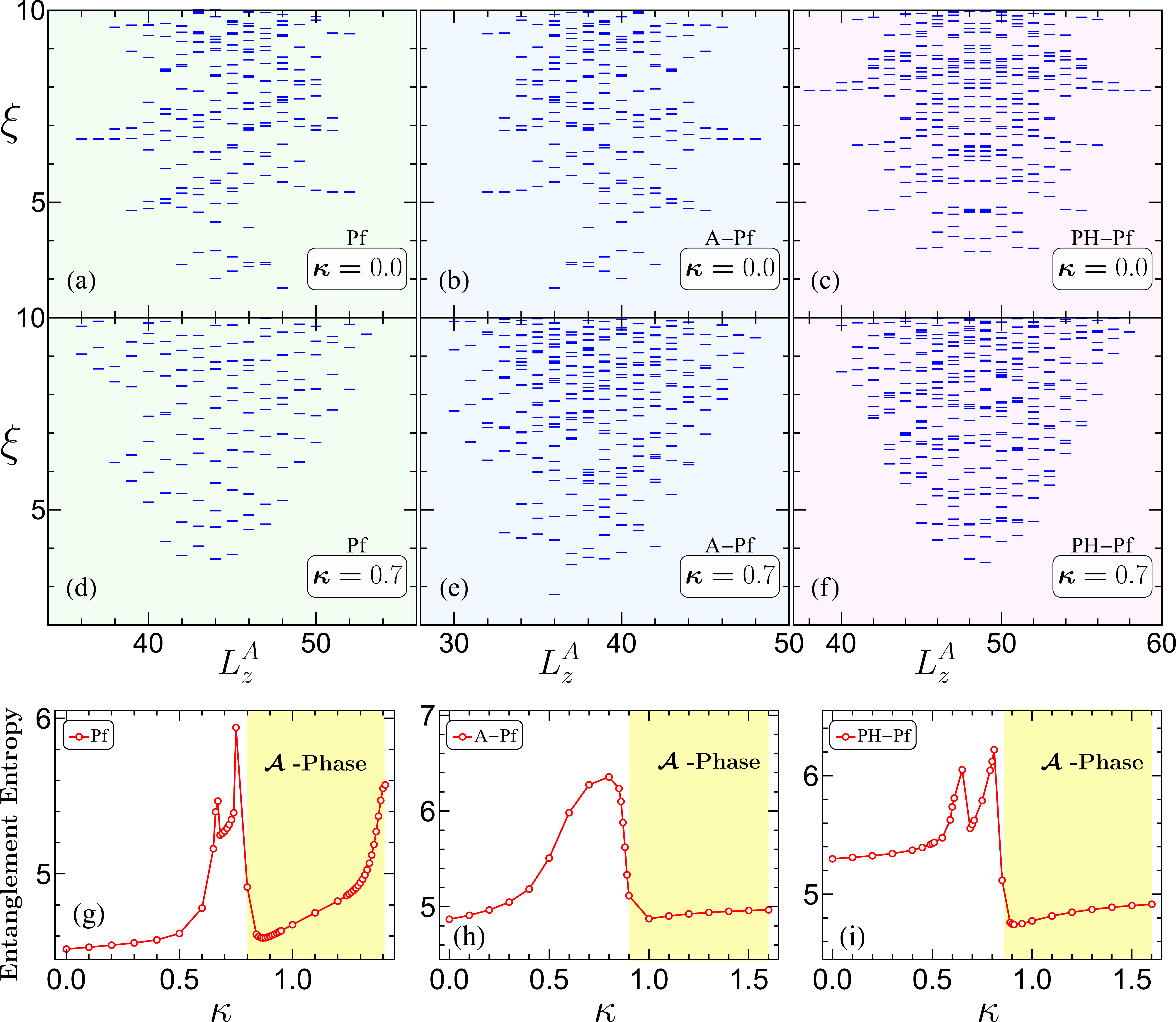}
	\caption{(a) ES for Pf flux shift with $ N_\phi = 25 $ at a $ \kappa $ belonging to low-$ \kappa $ phase. $ L^A_z $ represents the sum of the azimuthal components of angular momenta occupied by the particles in A part of the partition. Equal	number of electrons ($ N_A = N_B = 7 $) in both the partitions are considered for computing the corresponding ES. Here $ \xi $ represents
		entanglement energy in an arbitrary unit.
		(b) Same as (a) but for A-Pf flux shift with $ N_\phi = 29 $.
		(c) Same as (a) but for PH-Pf flux shift with $ N_\phi = 27 $.
		(d) ES for Pf flux shift ($ N_\phi = 25 $) at a $ \kappa $ belonging to unquantized phase.
		(e) Same as (d) but for A-Pf flux shift ($ N_\phi = 29 $).
		(f) Same as (d) but for PH-Pf flux shift ($ N_\phi = 27 $).
		(g)	Entanglement entropy versus $ \kappa $ for Pf flux shift ($ N_\phi = 25 $).
		(h) Entanglement entropy versus $ \kappa $ for A-Pf flux shift ($ N_\phi = 29 $).
		(i) Entanglement entropy versus $ \kappa $ for PH-Pf flux shift ($ N_\phi = 27 $).
		The shaded regions in (g)-(i) denote the A-phase.
	}
	\label{fig.es_ee}
\end{figure}

Here the set of single-particle orbitals
$\lambda=\{-Q,-Q+1,...,Q-1,Q\}$ is divided into two disjoint sets $\mu=\{-Q,...,Q-l_A-1\}$ and
$\nu=\{Q-l_A,...Q-1,Q\}$  respectively for block-\textit{B} and block-\textit{A}. If there are $N_A$ and $N_B$ number of electrons in the block-\textit{A} and -{\it B} respectively, one finds the corresponding basis states $|\nu_i\rangle$ and $|\mu_i\rangle$ while the basis states for the whole system are denoted by $|\lambda_i\rangle$.
%
Therefore, the ground state (or the trial state) may be expressed as
\begin{equation}
	|\psi\rangle = \sum_{\{\lambda_i\}} c_{\lambda_i} |\lambda_i\rangle = \sum_{\{\mu_i\},\{\nu_j\}} (O_f)_{ij} \text{ } |\mu_i\rangle \otimes |\nu_j\rangle \, ,
\end{equation}
where the component of entanglement matrix $O_f$ is the coefficient of $|\mu_i,\nu_j\rangle$ basis state of $\psi$, i.e., $(O_f)_{ij}= c_{(\mu_i,\nu_j)}$.
This $O_f$ matrix is block diagonal where each block is labeled by $ N_A $ and $L_z^A= \sum_{i=1}^{N_A}\nu_i$.


While the reduced density matrix for the \textit{A}-block may be expressed as $\rho_A= O_f O_f^\dagger$, it is $\rho_B= O_f^\dagger O_f$ for the block-\textit{B}. Consequently, EE of the state $|\psi\rangle$ is given by $S = S_A =  -\text{Tr}[\rho_A \text{ ln} \rho_A] = -\text{Tr}[\rho_B \text{ ln} \rho_B] = S_B$.

For a given $N_A$ and $L_z^A$, the singular value decomposition of the matrix $O_f$ will give rise to
\begin{eqnarray}
	|\psi \rangle &=&\sum_{\{\mu_i\},\{\nu_j\}} (O_f)_{ij} \text{ } |\mu_i\rangle \otimes |\nu_j\rangle	\nonumber \\
	&=& \sum_{i=1}^{\text{Rank}(O_f)} e^{-\xi_i/2} |\psi_B^i\rangle \otimes |\psi^i_A\rangle \, ,
\end{eqnarray}
where $e^{-(1/2)\xi_i} \ge 0, \quad |\psi_A^i\rangle \in A \text{ block }, \quad |\psi^i_B\rangle \in B \text{ block }$ and $\langle\psi_A^i|\psi_A^j\rangle = \langle\psi_B^i|\psi_B^j\rangle = \delta_{ij}$.
$\xi_i$ and $|\psi^i_A\rangle$ (or $|\psi^i_B\rangle$) respectively are the eigenvalues and the eigenstates of the reduced density matrix $\rho_A$ (or $\rho_B$) of the considered block.
If we construct a fictitious Hamiltonian $H_A$ using reduced density matrix  $\rho_A$ such that $H_A = - \text{log }\rho_A$, then, $\xi_i$ and $|\psi^i_A\rangle$ are the eigenvalues and eigenstates of this Hamiltonian. 
In the calculation of this ES with eigen values $\xi_i$ for orbital partitioning, we have considered the  orbitals in two blocks given in sets $\mu$ and $\nu$ with $l_A=(2Q
+1)/2$, and equal number of particles $(N_A = N/2 = N_B)$ in both the blocks.

In Fig.~S\ref{fig.es_ee}(a)-(c) ES is shown at Pf flux shift of $ N_\phi=25 $, A-Pf flux shift of $ N_\phi=29 $, and PH-Pf flux shift of $ N_\phi=27 $ respectively at the Coulomb limit i.e. $ \kappa=0.0 $ and well gapped ES can be seen for all those cases as expected. But there was no gap observed in the ES in the unquantized regime of $ \kappa=0.7 $ for the Pf, A-Pf, and PH-Pf flux shifts, Fig.~S\ref{fig.es_ee}(d)-(f). The Entanglement entropy with the variation of $ \kappa $ for the corresponding Pf, A-Pf, and PH-Pf flux shifts are shown in Fig.~S\ref{fig.es_ee}(g)-(i). A peak is observed in the entanglement entropy when the system undergoes the phase transition through the unquantized regime at around $ \kappa \simeq 0.7 $ and again attains a stable value in the $ \mathcal{A} $-phase.


\section{\label{sec.ovlp psiA difff flux}Overlap of $\Psi_{\mathcal{A}} $ with the exact ground states for three different flux-shifts}
For a fractional quantum Hall system in spherical geometry with a given number of flux quanta $ N_\phi =2Q $ where $Q$ is the magnetic monopole charge placed at the center of the sphere, the $ z $-component of orbital angular momentum is a good quantum number and the allowed values  (in the lowest Landau level) are,
\begin{equation}
	l_z = -Q,-(Q-1),...,Q-1,Q
\end{equation}
with degeneracy $2Q+1$.
$ N < 2Q+1$ electrons can occupy such quantum states in several possible ways with the criteria that the sum of the occupied $l_z$  must be zero $ ( \sum_{i=1}^N l^i_z = 0 ) $ if we fix the $z$-component of total angular momentum, $L_z$ to be zero. We consider $L_z =0$ so that we do not lose any angular momentum $L$. 
Now, for another system with  $ Q^\prime > Q $, such that $Q'=Q+q$ ($q$ integer),  $z$-components of angular momenta span as,
\begin{equation}
	l_z^\prime = -Q^\prime,-(Q^\prime-1),....,Q^\prime-1,Q^\prime
\end{equation}
Therefore, $N$-particle Hilbert space for $Q$ is a subspace of the same for $Q'$.
As an example, for $ N=4 $ particles the single-particle eigen basis at Pf flux are, \{$-$2.5, $-$1.5, 1.5, 2.5\}, \{$-$2.5, $-$0.5, 0.5, 2.5\}, \{$-$1.5, $-$0.5, 0.5, 1.5\}
and at PH-Pf flux are,
\{$-$3.5, $-$2.5, 2.5, 3.5\},  
\{$-$3.5, $-$1.5, 1.5, 3.5\}, 
\{$-$3.5, $-$0.5, 0.5, 3.5\}, 
\{$-$2.5, $-$1.5, 0.5, 3.5\}, 
\{$-$3.5, $-$0.5, 1.5, 2.5\}, 
\{$-$2.5, $-$1.5, 1.5, 2.5\}, 
\{$-$2.5, $-$0.5, 0.5, 2.5\}, 
\{$-$1.5, $-$0.5, 0.5, 1.5\}.
Clearly, the Hilbert space for the Pf shift is the subspace of the Hilbert space of the PH-Pf shift, and similarly it can be shown that the Hilbert space for the PH-Pf shift is the subspace of the Hilbert space of the A-Pf shift. 
Although our proposed wave function in Eq.~(2) corresponds to the PH-Pf flux shift, our hunch is that it will possess sizable overlap in the other two Hilbert spaces of the Pf and A-Pf as well, as the ES in the $\mathcal{A}$-phase is independent of these shifts.
We calculate the overlap of $\Psi_\mathcal{A}$ in Eq.~(2) for 14 particles with the exact ground state at the above three flux shifts for different values of $\kappa$ using the Monte Carlo method. In the low-$ \kappa $ phase which is identified as the Pf or A-Pf phase, the overlap, as expected, is negligibly small (Fig.~S\ref{fig.overlap}) for all the three flux shifts. On the other hand, $\Psi_{\cal A}$ in Eq.~(2) not only has a high overlap (Fig.~S\ref{fig.overlap}) in the $\mathcal{A}$-phase with the PH-Pf shift, it has an appreciable ($\sim 40-50\%$) overlap for the Pf and A-Pf shifts as well. This is quite unusual because $\Psi_{\mathcal{A}}$ corresponds to a flux shift that neither agrees with the Pf nor A-Pf shifts, although they share some part of the Hilbert space. The significant overlaps between two ground states at different flux shifts occur because the corresponding  weight factors of Fock-space are correlated in sign for the common Hilbert sub-space.

\begin{figure}[]
	\centering
	\includegraphics[width=\linewidth]{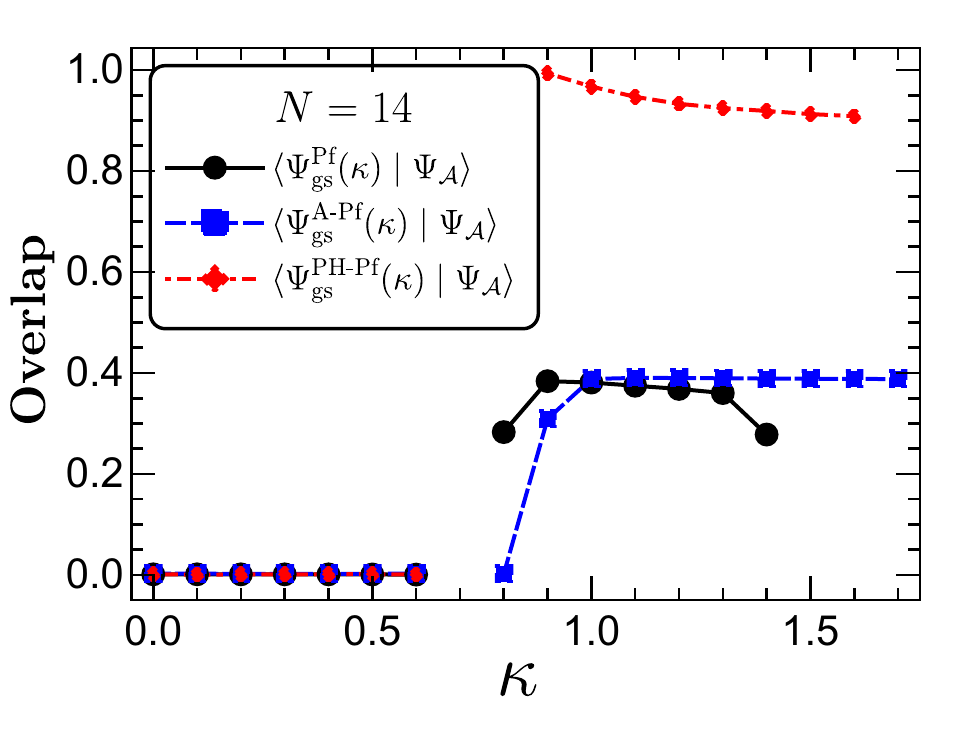}
	\caption{(color online)
		Overlap of $ \Psi_{\mathcal{A}} $ in Eq.~(2) shown for $ N=14 $ particles with the exact ground states at Pf flux$ =25 $, A-Pf flux$ =29 $, and PH-Pf flux$ =27 $ with varying Landau level mixing strength, $ \kappa $.}
	\label{fig.overlap}
\end{figure}


In contrary to the above findings, the overlap of the ground states at two different filling factors in the lowest Landau level at two different flux shifts but with same number of electrons is extremely low as shown in Table~S\ref{tab.lll_ovlp}, despite the sharing of common single-particle basis states as described above. As these two quantum Hall states are of different topologies, the weight factors of the common Hilbert subspace have uncorrelated signs and thus producing low overlap between the two states. 

\begin{table}[h]
	\begin{ruledtabular}
		\begin{tabular}{c|ccc}
			$ \quad N \quad$
			& $ \nu, N_{\phi} $
			& $ \nu^\prime, N_{\phi}^\prime $
			& $ \langle \Psi_{\text{gs}}^{(\nu, N_{\phi})} \mid \Psi_{\text{gs}}^{(\nu^\prime, N_{\phi}^\prime)} \rangle $                      \\
			\hline
			8  & 2/5, 16                                                                                                      & 2/3, 12 & 0.060856 \\
			10 & 2/5, 21                                                                                                      & 2/3, 15 & 0.002706 \\
			10 & 1/3, 27                                                                                                      & 2/5, 21 & 0.000213
		\end{tabular}
	\end{ruledtabular}
	\caption{Overlap of exact ground states of different fluxes but same number of particles in the lowest Landau level.}
	\label{tab.lll_ovlp}
\end{table}


\section{\label{sec.ovlp psiA PHPF}Overlap of $ \Psi_{\mathcal{A}} $ with the variation of $ \kappa $ at PH-Pf flux}
The overlaps of $ \Psi_{\mathcal{A}} $ in Eq.~(2) with the exact ground state corresponding to PH-Pf flux by varying $ \kappa $ for different number of particle systems are tabulated in Table~S\ref{tab.overlap}. The range of ${\cal A}$-phase appears to be increasing with $N$.
\begin{table}[h]
	\begin{ruledtabular}
		\begin{tabular}{cccccc}
			& \multicolumn{5}{@{}c@{}}{$ N $}                                              \\
			\cline{2-6}
			\scalebox{1.2}{$\quad \kappa \quad $} & 8                               & 10      & 12        & 14        & 16       \\
			\hline
			
			0.0                                   & 0.09797                         & -       & 0.0004(1) & 0.0010(2) & -        \\ 
			0.1                                   & 0.10267                         & -       & 0.0004(1) & 0.0006(3) & -        \\ 
			0.2                                   & 0.10929                         & -       & 0.0005(1) & 0.0009(3) & -        \\ 
			0.3                                   & 0.11933                         & -       & 0.0006(3) & 0.0010(3) & -        \\ 
			0.4                                   & 0.13652                         & -       & 0.0007(2) & 0.0012(5) & -        \\ 
			0.5                                   & 0.17442                         & -       & 0.0009(2) & 0.0014(6) & -        \\ 
			0.6                                   & -                               & -       & 0.0015(2) & 0.0006(2) & 0.003(1) \\ 
			0.7                                   & 0.95379                         & -       & 0.0028(2) & -         & -        \\ 
			0.8                                   & 0.99029                         & 0.99397 & 0.9761(1) & -         & -        \\ 
			0.9                                   & 0.96157                         & 0.96463 & 0.9778(1) & 0.9935(0) & 0.37(1)  \\ 
			1.0                                   & 0.93345                         & 0.94240 & 0.9517(1) & 0.9673(3) & 0.989(0) \\ 
			1.1                                   & 0.90923                         & 0.92629 & 0.9349(2) & 0.9470(2) & 0.966(1) \\ 
			1.2                                   & -                               & 0.91403 & 0.9232(2) & 0.9331(2) & 0.950(1) \\ 
			1.3                                   & -                               & 0.90429 & 0.9147(2) & 0.9247(4) & 0.939(1) \\ 
			1.4                                   & -                               & -       & 0.9082(3) & 0.9190(5) & 0.932(1) \\ 
			1.5                                   & -                               & -       & -         & 0.9128(2) & 0.926(1) \\ 
			1.6                                   & -                               & -       & -         & 0.9087(5) & 0.922(1) \\ 
			1.7                                   & -                               & -       & -         & -         & 0.919(1) \\ 
			1.8                                   & -                               & -       & -         & -         & -        \\ 
			1.9                                   & -                               & -       & -         & -         & -        \\
		\end{tabular}
	\end{ruledtabular}
	\caption{
		Overlaps of  $\Psi_{\mathcal{A}}$ in Eq.~(2) with the exact ground states of $\hat{H}_{\text{eff}}$  in Eq.~(1) at PH-Pf shift for different values of $\kappa$ and for $ N=8$, 10, 12, 14 and 16 electrons. The dashed mark shells represent unquantized phase between two quantized phases: low-$\kappa$ conventional phase and the ${\cal A}$-phase at moderate-$\kappa$. There is no low-$\kappa$ phase at PH-Pf flux for $N=10$ and 16. The numbers within parentheses indicate uncertainty in the Monte Carlo evaluation. For $ N=8 $ and $ 10 $ we get the exact overlap value through decomposition [S\onlinecite{das_2021_UnconventionalFillingFactor}] into single particle eigen basis. }
	\label{tab.overlap}
\end{table}

%

\end{document}